

\font\twelverm=cmr10 scaled 1200    \font\twelvei=cmmi10 scaled 1200
\font\twelvesy=cmsy10 scaled 1200   \font\twelveex=cmex10 scaled 1200
\font\twelvebf=cmbx10 scaled 1200   \font\twelvesl=cmsl10 scaled 1200
\font\twelvett=cmtt10 scaled 1200   \font\twelveit=cmti10 scaled 1200
\font\twelvesc=cmcsc10 scaled 1200

\skewchar\twelvei='177   \skewchar\twelvesy='60

\def\twelvepoint{\normalbaselineskip=12.4pt
  \abovedisplayskip 12.4pt plus 3pt minus 6pt
  \belowdisplayskip 12.4pt plus 3pt minus 6pt
  \abovedisplayshortskip 0pt plus 3pt
  \belowdisplayshortskip 7.2pt plus 3pt minus 4pt
  \smallskipamount=3.6pt plus1.2pt minus1.2pt
  \medskipamount=7.2pt plus2.4pt minus2.4pt
  \bigskipamount=14.4pt plus4.8pt minus4.8pt
  \def\rm{\fam0\twelverm}          \def\it{\fam\itfam\twelveit}%
  \def\sl{\fam\slfam\twelvesl}     \def\bf{\fam\bffam\twelvebf}%
  \def\mit{\fam 1}                 \def\cal{\fam 2}%
  \def\tt{\twelvett}
  \def\sc{\twelvesc}
  \def\nullspace{\nulldelimiterspace=0pt \mathsurround=0pt }
  \def\big##1{{\hbox{$\left##1\vbox to 10.2pt{}\right.\nullspace$}}}
  \def\Big##1{{\hbox{$\left##1\vbox to 13.8pt{}\right.\nullspace$}}}
  \def\bigg##1{{\hbox{$\left##1\vbox to 17.4pt{}\right.\nullspace$}}}
  \def\Bigg##1{{\hbox{$\left##1\vbox to 21.0pt{}\right.\nullspace$}}}
  \textfont0=\twelverm   \scriptfont0=\tenrm   \scriptscriptfont0=\sevenrm
  \textfont1=\twelvei    \scriptfont1=\teni    \scriptscriptfont1=\seveni
  \textfont2=\twelvesy   \scriptfont2=\tensy   \scriptscriptfont2=\sevensy
  \textfont3=\twelveex   \scriptfont3=\twelveex  \scriptscriptfont3=\twelveex
  \textfont\itfam=\twelveit
  \textfont\slfam=\twelvesl
  \textfont\bffam=\twelvebf \scriptfont\bffam=\tenbf
  \scriptscriptfont\bffam=\sevenbf
  \normalbaselines\rm}

\def\beginlinemode{\endmode
  \begingroup\parskip=0pt \obeylines\def\\{\par}\def\endmode{\par\endgroup}}
\def\beginparmode{\endmode
  \begingroup \def\endmode{\par\endgroup}}
\let\endmode=\par
{\obeylines\gdef\
{}}
\def\singlespace{\baselineskip=\normalbaselineskip}

\def\oneandahalfspace{\baselineskip=\normalbaselineskip
  \multiply\baselineskip by 3 \divide\baselineskip by 2}
\def\doublespace{\baselineskip=\normalbaselineskip \multiply\baselineskip by 2}

\newcount\firstpageno
\firstpageno=2
\footline={\ifnum\pageno<\firstpageno{\hfil}\else{\hfil\twelverm\folio\hfil}\fi}
\let\rawfootnote=\footnote		
\def\footnote#1#2{{\rm\singlespace\hang
  \rawfootnote{#1}{#2\hfill\vrule height 0pt depth 6pt width 0pt}}}
\def\smallfootnote#1#2{{\singlespace\hang
  \rawfootnote{#1}{\tenrm#2\hfill\vrule
                                  height 0pt depth 6pt width 0pt}}}
\def\raggedcenter{\leftskip=4em plus 12em \rightskip=\leftskip
  \parfillskip=0pt \spaceskip=.3333em \xspaceskip=.5em
  \pretolerance=9999 \tolerance=9999
  \hyphenpenalty=9999 \exhyphenpenalty=9999 }
\def\dateline{\rightline{\ifcase\month\or
  January\or February\or March\or April\or May\or June\or
  July\or August\or September\or October\or November\or December\fi
  \space\number\year}}
\def\received{\vskip 3pt plus 0.2fill
 \centerline{\sl (Received\space\ifcase\month\or
  January\or February\or March\or April\or May\or June\or
  July\or August\or September\or October\or November\or December\fi
  \qquad, \number\year)}}

\hsize=6.5truein
\vsize=8.9truein
\parskip=\medskipamount
\twelvepoint		
\overfullrule=0pt	

\def\preprintno#1{
 \rightline{\rm #1}}	
\def\title			
  {\null\vskip 3pt plus 0.3fill \beginlinemode
   \doublespace \raggedcenter \bf}
\def\author			
  {\vskip 3pt plus 0.3fill \beginparmode \raggedcenter \sc}
\def\and			
  {\vskip 3pt plus 0.3fill \raggedcenter \rm and}
\def\affil			
  {\vskip 3pt plus 0.1fill \beginlinemode
   \oneandahalfspace \raggedcenter \sl}
\def\abstract			
  {\vskip 3pt plus 0.3fill \beginparmode
   \oneandahalfspace \narrower ABSTRACT:~~}
\def\endtitlepage{\vfill\eject\body}
\def\body{\beginparmode}
\def\subhead#1{\vskip 0.25truein{\raggedcenter #1 \par}
   \nobreak\vskip 0.25truein\nobreak}
\def\references
  {\subhead{REFERENCES}
   \frenchspacing \parindent=0pt \leftskip=0.8truecm \rightskip=0truecm
   \parskip=4pt plus 2pt \everypar{\hangindent=\parindent}}
\def\refstylenp{		
  \gdef\refto##1{~[##1]}				
  \gdef\r##1{~[##1]}	         			
  \gdef\refis##1{\indent\hbox to 0pt{\hss[##1]~}}     	
  \gdef\citerange##1##2##3{~[\cite{##1}--\setbox0=\hbox{\cite{##2}}\cite{##3}]}
  \gdef\journal##1, ##2, ##3,                           
    ##4{{\sl##1} {\bf ##2} (##3) ##4}
  \gdef\eq{eq.}
  \gdef\eqs{eqs.}
  \gdef\Eq{Eq.}
  \gdef\Eqs{Eqs.} }

\def\prd{\journal Phys. Rev. D}
\def\prl{\journal Phys. Rev. Lett.}

\def\np{\journal Nucl. Phys.}
\def\pl{\journal Phys. Lett.}

\def\endreferences{\body}

\def\endit{\endmode\vfill\supereject\end}
\def\frac#1#2{{\textstyle{#1 \over #2}}}
\def\half{{\textstyle{ 1\over 2}}}

\def\sss{\scriptscriptstyle}

\def\twiddle{\lower.9ex\rlap{$\kern-.1em\scriptstyle\sim$}}
\def\bigtwiddle{\lower1.ex\rlap{$\sim$}}
\def\gtwid{\mathrel{\raise.3ex\hbox{$>$\kern-.75em\lower1ex\hbox{$\sim$}}}}
\def\ltwid{\mathrel{\raise.3ex\hbox{$<$\kern-.75em\lower1ex\hbox{$\sim$}}}}
\def\square{\kern1pt\vbox{\hrule height 1.2pt\hbox{\vrule width 1.2pt\hskip 3pt
   \vbox{\vskip 6pt}\hskip 3pt\vrule width 0.6pt}\hrule height 0.6pt}\kern1pt}
\def\ucsb{Department of Physics\\University of California\\
Santa Barbara, CA 93106}

\def\p{\varphi}

\def\r{\rho}

\def\O{\Omega}

\def\tev{{\rm \,Te\kern-0.125em V}}
\def\gev{{\rm \,Ge\kern-0.125em V}}
\def\mev{{\rm \,Me\kern-0.125em V}}
\def\kev{{\rm \,ke\kern-0.125em V}}
\def\ev{{\rm \,e\kern-0.125em V}}

\def\sla{\raise.15ex\hbox{$/$}\kern-.57em}
\refstylenp

\catcode`@=11
\newcount\r@fcount \r@fcount=0
\newcount\r@fcurr
\immediate\newwrite\reffile
\newif\ifr@ffile\r@ffilefalse
\def\w@rnwrite#1{\ifr@ffile\immediate\write\reffile{#1}\fi\message{#1}}
\def\writer@f#1>>{}
\def\referencefile{
  \r@ffiletrue\immediate\openout\reffile=\jobname.ref%
  \def\writer@f##1>>{\ifr@ffile\immediate\write\reffile%
    {\noexpand\refis{##1} = \csname r@fnum##1\endcsname = %
     \expandafter\expandafter\expandafter\strip@t\expandafter%
     \meaning\csname r@ftext\csname r@fnum##1\endcsname\endcsname}\fi}%
  \def\strip@t##1>>{}}

\def\citeall#1{\xdef#1##1{#1{\noexpand\cite{##1}}}}
\def\cite#1{\each@rg\citer@nge{#1}}	
\def\each@rg#1#2{{\let\thecsname=#1\expandafter\first@rg#2,\end,}}
\def\first@rg#1,{\thecsname{#1}\apply@rg}	
\def\apply@rg#1,{\ifx\end#1\let\next=\relax
\else,\thecsname{#1}\let\next=\apply@rg\fi\next}
\def\citer@nge#1{\citedor@nge#1-\end-}	
\def\citer@ngeat#1\end-{#1}
\def\citedor@nge#1-#2-{\ifx\end#2\r@featspace#1 
  \else\citel@@p{#1}{#2}\citer@ngeat\fi}	
\def\citel@@p#1#2{\ifnum#1>#2{\errmessage{Reference range #1-#2\space is bad.}%
    \errhelp{If you cite a series of references by the notation M-N, then M and
    N must be integers, and N must be greater than or equal to M.}}\else%
 {\count0=#1\count1=#2\advance\count1
by1\relax\expandafter\r@fcite\the\count0,%
  \loop\advance\count0 by1\relax
    \ifnum\count0<\count1,\expandafter\r@fcite\the\count0,%
  \repeat}\fi}
\def\r@featspace#1#2 {\r@fcite#1#2,}	
\def\r@fcite#1,{\ifuncit@d{#1}
    \newr@f{#1}%
    \expandafter\gdef\csname r@ftext\number\r@fcount\endcsname%
                     {\message{Reference #1 to be supplied.}%
                      \writer@f#1>>#1 to be supplied.\par}%
 \fi%
 \csname r@fnum#1\endcsname}
\def\ifuncit@d#1{\expandafter\ifx\csname r@fnum#1\endcsname\relax}%
\def\newr@f#1{\global\advance\r@fcount by1%
    \expandafter\xdef\csname r@fnum#1\endcsname{\number\r@fcount}}
\let\r@fis=\refis			
\def\refis#1#2#3\par{\ifuncit@d{#1}
   \newr@f{#1}%
   \w@rnwrite{Reference #1=\number\r@fcount\space is not cited up to now.}\fi%
  \expandafter\gdef\csname r@ftext\csname r@fnum#1\endcsname\endcsname%
  {\writer@f#1>>#2#3\par}}
\def\ignoreuncited{
   \def\refis##1##2##3\par{\ifuncit@d{##1}%
     \else\expandafter\gdef\csname r@ftext\csname
r@fnum##1\endcsname\endcsname%
     {\writer@f##1>>##2##3\par}\fi}}
\def\r@ferr{\endreferences\errmessage{I was expecting to see
\noexpand\endreferences before now;  I have inserted it here.}}
\let\r@ferences=\references
\def\references{\r@ferences\def\endmode{\r@ferr\par\endgroup}}
\let\endr@ferences=\endreferences
\def\endreferences{\r@fcurr=0
  {\loop\ifnum\r@fcurr<\r@fcount
    \advance\r@fcurr by 1\relax\expandafter\r@fis\expandafter{\number\r@fcurr}%
    \csname r@ftext\number\r@fcurr\endcsname%
  \repeat}\gdef\r@ferr{}\endr@ferences}

\def\range#1#2#3{\citerange{#1}{#2}{#3}}
\let\r@fend=\endpaper\gdef\endpaper{\ifr@ffile
\immediate\write16{Cross References written on []\jobname.REF.}\fi\r@fend}
\catcode`@=12
\citeall\refto		
\citeall\r		%

\ignoreuncited

\def\mpl{m_{\rm\sss Pl}}
\def\Oh2{\O h^2}

\singlespace
\preprintno{UCSBTH--92--05}
\preprintno{April 1992}

\title CHAOTIC DARK MATTER

\author Raghavan Rangarajan\smallfootnote{$^{\rm (a)}$}
{Electronic address (internet): raghu@pcs3.ucsb.edu.}

\and

\author Mark Srednicki\smallfootnote{$^{\rm (b)}$}
{Electronic address (internet): mark@tpau.physics.ucsb.edu.}
\affil\ucsb

\abstract
A very weakly coupled scalar field with mass $m$ and initial vacuum
expectation value $V$ will provide enough mass to close the universe
provided $V\simeq (3\times 10^8\gev)(100\gev/m)^{1/4}$.
We discuss possible models in which such a field could arise.


\endtitlepage
\body
\oneandahalfspace

Particle physics has provided a number of plausible dark matter candidates,
most notably massive but light neutrinos\r{gz66},
the lightest supersymmetric particle\range{goldberg83}{krauss83a}{ehnos84},
and the axion\range{pww83}{as83}{df83}.
These last two would be cold dark matter, the most plausible type.
(For a review of dark matter, see Ref.\r{srednicki90}.)
The axion is especially interesting;
a closure density of axions does not arise from thermal processes
(as would be expected in a hot early universe), but rather from a coherent
state of particles which approximate a classical field with a definite
value.  Thermal effects do play an important role, however.
At very high temperatures, the axion field has no potential energy, and
so assumes an arbitrary initial vacuum expectation value (VEV) $V$,
determined by the
details of the initial conditions.  In most models of early cosmology,
this arbitrary value is essentially random and unpredictable; all we know is
that it lies on a circle of circumference $2\pi f$, where $f$ is the
axion decay constant, related to the axion mass $m_a$
via $m_a\sim(100\mev)^2/f$.  As the
temperature drops, however, a potential turns on
(proportional to $T^{-8}$), and eventually the
axion field begins to oscillate about the minimum determined by the potential.
The contribution $\O$ of these coherent axions to the mass density
of the universe (in units of the critical density) is given
by\range{pww83}{as83}{df83}
$$\Oh2 \sim 0.3 \left({V\over 10^{12}\gev}\right)^2
                \left({m_a\over 10^{-5}\ev}\right),             \eqno(1)$$
where $h$ is the current value of the Hubble parameter in units of
$100{\rm\,km/s\!\cdot\!Mpc}$.
(For a more detailed analysis, see Ref.\r{turner86}.)
Because the initial axion VEV $V$ lies on a circle of radius $f$,
it is likely that $V\sim f$.  Putting this into \eq(1) and using the relation
between $f$ and $m_a$ gives an upper limit on $f$ of about $10^{12}\gev$.

In this paper we examine the case of a scalar field
with mass $m$ and VEV $V$ which we imagine to be so weakly coupled that
thermal effects can be neglected entirely.
We find that the contribution of such a field to the mass density of the
universe is
$$\Oh2\sim 0.7\left({V\over 3\times10^8\gev}\right)^2
              \left({m\over 100\gev}\right)^{1/2}.         \eqno(2)$$
Note that $\O\sim V^2 m^{1/2}$, in contrast to the $V^2 m$ dependence in the
case of the axion.  While this formula is easy to derive, to our knowledge
it has never appeared in the literature.

Are such fields likely to occur?  We note that string theory (with four
uncompactified dimensions) always has at least four massless scalar
particles (the string dilaton, the breathing mode of the compact dimensions,
and two partners demanded by supersymmetry)\r{witten85}.
These particles can get mass
only after supersymmetry is broken.  It is widely anticipated that the
spontaneous breakdown of SU(2)$\times$U(1) will turn out to be related
to supersymmetry breaking, yielding natural masses of order the weak
interaction scale, $\sim\!\!100\gev$, for these scalars.
However, there is no reason to believe
that these scalars will emerge from the early universe (after a period of
inflation, say) at the minima of the potentials generated by supersymmetry
breaking.  They may have large, arbitrary VEVs.

Indeed, a very similar proposal has been made by Linde\r{linde83},
but to provide inflation rather than dark matter.
Linde envisions a region of space in which a scalar field with
mass $m$ happens to have a large VEV $V$, large enough that the stored
energy density dominates over that of all other particles and fields.
This energy density drives inflation.  Linde refers to this as
``chaotic inflation,'' since the field takes on a chaotic pattern of values
throughout the universe.  We envision a similar scenario,
except that we assume the energy density stored in our field happens to
be much less than the thermal energy of ambient hot particles, which we
imagine were produced by reheating after a period of inflation.
Our field does not drive inflation, but it does provide for dark matter.
Given the similarity to chaotic inflation, we refer to this scenario
as chaotic dark matter.

Of course, to be a dark matter candidate our particles must be
stable or nearly so,
with a decay rate no larger than the Hubble parameter.  Also, they
must be self-coupled weakly enough to avoid efficient transfer of energy out
of the coherent state.  As we will see,
this latter constraint is not a problem, but
the former is an impediment to identifying our field as the dilaton
or one of its cousins.  We would generically expect a dilaton with mass~$m$
to have a decay rate of order $m^3/\mpl^2$, where $\mpl=1.2\times10^{19}\gev$
is the Planck mass.  (This is because
the dilaton should couple with gravitational strength to all other fields.)
Requiring this to be less than the Hubble parameter
translates into $m\ltwid 100\mev$, a factor of a thousand smaller than
the weak interaction scale.  We will not attempt to solve this dilemma here;
like the field which is hypothesized to drive chaotic inflation,
our field need not be one of the string fields mentioned above.
Producing a closure density with a mass of $100\mev$ requires a VEV of
$2\times10^9\gev$.

We now turn to demonstrating \eq(2).  For the most part we can follow the
analysis originally used for the axion\range{pww83}{as83}{df83}.
The equation of motion for our field $\p$ is
$$\ddot\p-R^{-2}\nabla^2\p+3(\dot R/R)\dot\p+m^2\p=0,       \eqno(3)$$
where $R(t)$ is the scale factor; $R\sim t^{1/2}$ when the universe is
radiation dominated and $R\sim t^{2/3}$ when the universe is matter dominated.
We take the initial value of $\p(\vec x,t)$ to be $V$ everywhere; possible
spatial inhomogeneities will be treated later.
Until the age $t$ of the universe is comparable to the period $1/m$ of the
oscillations, the field $\p$ will not have had enough time to change
significantly, and hence remains frozen at $\p(t)=V$.
Once $t$ becomes of order $1/m$, however, $\p(t)$ begins to oscillate.
Call the temperature at this time $T_1$.
Afterward, the time-averaged
energy density in the field is $\rho=\half m^2\p^2$,
with $\p(t)=V$ when $T=T_1$.
This energy density decreases
like the energy density of particles with zero momentum; that is,
it decreases as $1/R(t)^3$,
or equivalently as $T^3$, where $T$ is the temperature.
We thus find
$$\rho=\half m^2V^2(T_0/T_1)^3,       \eqno(4)$$
where $T_0=2.75\,$K is the present temperature.  We compute $T_1$ from
the usual time-temperature relation, $t=\mpl/(\gamma T^2)$, where $\gamma=34.3$
in the standard model and $\gamma=49.4$ in the supersymmetric standard model.
Using $t=1/m$ and $\gamma=34.3$ we find
$$T_1\sim(6\times10^9\gev)\left(m\over100\gev\right)^{1/2}.   \eqno(5) $$
Putting this into \eq(4) and dividing by the critical density
$\rho_{\rm c}=3H^2/8\pi G$ yields \eq(2), our central result.

We shall now consider self interactions of the $\p$ field.  In keeping with
the idea that its potential is generated by supersymmetry breaking,
we parameterize the interactions as
$${\cal L}_{\rm int}(\p)=-m^2\sum_{n=3}^\infty \lambda_n\,\mpl^{-(n-2)}\p^n
                                                              \eqno(6)$$
where the $\lambda_n$ are dimensionless numbers, assumed to be
of order one.  (Derivative interactions could also be included,
and would lead to similar results.)
These interactions will allow energy to be transferred out of the zero-momentum
particles and into particles with relativistic energies; the energy will then
red-shift away faster, like $T^4$ instead of $T^3$, for as long as the
particles remain relativistic.

Following the analysis of Ref.\r{as83} for the axion,
we consider a perturbation of the form
$$ \p(\vec x,t)=A(t)\cos(mt) +  A_k(t)e^{i\vec k\cdot\vec x}       \eqno(7) $$
where $A_k$ is treated as infinitesimal.  Substituting in
\eq(3) and taking $R$ and $A$ to be approximately constant, we get
$$\ddot A_k(t)+\bigl[(k/R)^2+m^2+6\lambda_3m^2(A/\mpl)\cos(mt)
                                     + \ldots\bigr]A_k(t) =0,      \eqno(8) $$
where the ellipses stand for the higher order terms in \eq(6).  These will
be suppressed by additional powers of $A/\mpl$, and can be neglected.
\Eq(8) is Mathieu's equation.  For certain bands of values of $k$,
$A_k$ grows exponentially.  This would mean that there is an instability
which could pump energy into high momentum modes.  However, as noted above,
we need only be concerned with modes for which $k\gg m$, and these modes
account
for an exponentially small range of values of $k$ at large $k$.
Thus the unstable
modes do not have enough time to grow significantly before they are redshifted
into the regime of stable wavenumbers.  Therefore,
just as for the axion, this effect can be ignored.

To conclude, we have shown that a coherent scalar field with mass $m$
and initial VEV $V$ contributes to the mass density of the universe according
to \eq(2).  A field with $m\sim100\gev$ and $V\sim3\times10^8\gev$ would
provide a critical density of cold dark matter.  Such a field might
arise is certain string models, but generically the corresponding particles
would have lifetimes that are too short unless the mass is reduced
to $m\ltwid 100\mev$.

This work was supported in part by NSF Grant No.~PHY-86-14185.

\references

\hyphenation{Amsterdam}

\refis{gz66}S. S. Gershtein and Ya. B. Zel'dovich,
\journal JETP Lett., 4, 1966, 174.

\refis{goldberg83}H. Goldberg, \prl, 50, 1983, 1419.

\refis{krauss83a}L. M. Krauss, \np, B227, 1983, 556. 

\refis{ehnos84}J. Ellis, J. S. Hagelin, D. V. Nanopoulos, K. A. Olive, and
M. Srednicki, \np, B238, 1984, 453.

\refis{pww83}J. Preskill, M. B. Wise, and F. Wilczek, \pl, B120, 1983, 127.

\refis{as83}L. Abbott and P. Sikivie, \pl, B120, 1983, 133.

\refis{df83}M. Dine and W. Fischler, \pl, B120, 1983, 137.

\refis{srednicki90}Particle Physics and Cosmology: Dark Matter,
ed. M. Srednicki (North--Holland, Amsterdam, 1990).

\refis{turner86}M. S. Turner, \prd, 33, 1986, 889.  

\refis{witten85}E. Witten, \pl, B155, 1985, 151. 

\refis{linde83}A. D. Linde, \pl, 129B, 1983, 177.

\endreferences\endit\end